# Characteristics of International versus Non-International Scientific Publication Media in Team- and Author-Based Data

Nadine Rons*

* *Nadine.Rons@vub.ac.be*
Research Coordination Unit, Vrije Universiteit Brussel (VUB), Pleinlaan 2, B-1050 Brussels (Belgium)

**Introduction**

The enlarged coverage of the international publication and citation databases Web of Science and Scopus towards local media in social sciences was a welcome response to an increased usage of these databases in evaluation and funding systems. The mostly international journals available earlier were the basis for the development of current standard bibliometric indicators. The same indicators may no longer measure exactly the same concepts when applied to newly introduced or extended media categories, with possibly different characteristics than those of international journals. This paper investigates differences between media with and without international dimension in publication data at team and author level. The findings relate the international publication categories to research quality, important for validation of their usage in evaluation or funding models that aim to stimulate quality.

**Data and Methodology**

*Team-based research assessment data:*
Publication productivity and peer ratings were collected per team from a cycle of research assessments by international expert panels, organized per discipline at the Vrije Universiteit Brussel (Rons et al., 2008) and completed in 2011:
- Publication productivity includes contributions in a five year period in books, journals and proceedings, each divided into subcategories with and without international dimension, and calculated per 'leading researcher' (level required for main promotership of research projects and PhD's) to normalize for team size. Each team validated its publication and personnel data.
- Peer ratings include an overall evaluation score and scores on scientific merit, planning, innovation, team quality, feasibility, productivity and scientific impact, on a scale from 1 to 10. Mechanisms used to limit the influence of an expert's bias or rating habits are described by Rons et al. (2008) and, in the context of comparisons to other indicators, by Rons & Spruyt (2006).

Correlations between productivity and peer ratings were calculated per publication category, for disciplines with at least five teams and where at least half of the teams publish in that category. Disciplines were grouped to calculate correlations for the broad domains of social sciences and humanities (D1), basic, applied and biomedical sciences (D2), and all disciplines combined. Before combining productivity values and peer ratings, both were normalized per discipline according to average values and standard deviations. This is an important step for the global analysis, as shown in an earlier application to a more limited set of disciplines and a different set of output categories (Rons & De Bruyn, 2007).

*Author-based Web of Science data:*
Article, letter and review type publications in the Social Science Citation Index ('SSCI-publications') were used to investigate differences between journal categories with and without international dimension, with English and non-English language as a proxy. Three appropriate combinations of a subject category and a country were selected for analysis, with:
- High percentage of non-English SSCI-publications (48-51%), to avoid low local journal coverage;
- Substantial volume of English SSCI-publications (94-155 per year), to avoid closed science systems.

For each combination, the associated sample was determined of authors who publish in both the English and the local language in the same time frame of five years (2006-2010), with:
- At least 1 English language, 1 non-English language and in total 3 SSCI-publications with the selected subject category and country affiliations;
- At least half of the SSCI-publications in the selected subject category;
- SSCI-publications in the first and the last year of the five year period.

For each author, data were collected separately for English ($e$) and non-English ($n$) SSCI-publications:
- Number of SSCI-publications: $Pe$, $Pn$;
- Highest number of citations obtained by a SSCI-publication: $Ce$, $Cn$;
- Percentage of uncited SSCI-publications: $Ue$, $Un$.

For each author sample, correlations between these parameters were calculated to investigate the relations between the English and non-English subcategories, and between these subcategories and citation levels as a proxy for quality.

## Results

In the team-based data (Table 1) higher productivity is found to correspond to higher peer esteem only for the international publication subcategories (shaded). Among the document types studied, publications in journals with international referee system are found to be the most important category with productivity reflecting research quality, followed by book chapters with international referee system and contributions in international conference proceedings.

Table 1. Team-based data: Significant correlations with peer ratings per scientific publication category

| Domain (Nr. of panels) | Publication category | | | | | | | | |
|---|---|---|---|---|---|---|---|---|---|
| | Book chapters (C) & journal articles (J) | | | | | | Communications at conferences | | |
| | Referee system | | | | | | Conf. scope | | |
| | Internat. | | Nat. | | None | | Internat. | | Other |
| | C | J | C | J | C | J | I | A | I | A |
| D1 (7) | | 8+ | 5- | 8- | 1- | 1- | | | 1- | |
| D2 (11) | 2+ | 8+ | | | | | 1+ | | | |
| Globally | 5+ | 8+ | | 1- | | 5- | 1+ | 2+ | | |

X+/-: For X peer ratings (out of 8) significantly positive (+) or negative (-) correlations are found with the publication category ($p \leq 0,05$).
I/A: Published integrally (I) or as abstract or not (A).

Table 2. Author-based data: Characteristics of English versus non-English publications

| Author sample | | Pearson product-moment correlations: $r$; $p$ ($\leq 0,05$ shaded) | | | |
|---|---|---|---|---|---|
| Country Subject Cat. | N | *Pe* vs. *Pn* | *Ce* vs. *Cn* | *Ue* vs. *Un* | *Ce* vs. *%e* |
| Germany Politic. Sc. | 23 | -0,16; 0,23 | -0,20; 0,18 | 0,17; 0,22 | 0,48; 0,01 |
| Spain Psychol.Md | 68 | -0,02; 0,44 | 0,11; 0,18 | 0,21; 0,04 | 0,26; 0,02 |
| Germany Social Sc. | 22 | -0,10; 0,33 | -0,25; 0,13 | 0,00; 0,49 | 0,45; 0,02 |

*Source: Web of Science (WoS), accessed online 04.03-05.04.2012. Data sourced from Thomson Reuters Web of Knowledge (formerly referred to as ISI Web of Science).*

In the author-based data (Table 2) productivities in the English and non-English subcategories (*Pe* vs. *Pn*) are found to be uncorrelated. For the two subcategories relations between highest citation levels (*Ce* vs. *Cn*) and between percentages uncited (*Ue* vs. *Un*) vary depending on the sample. For each of the three samples, higher numbers of citations obtained by English language publications correspond to higher percentages of English language publications in an author's record (*Ce* vs. *%e*, with *%e* = *Pe/(Pe+Pn)*).

## Discussion and Conclusions

The results demonstrate the different characteristics of international and non-international publication media, or English and non-English media as a proxy. Productivities in these two subcategories are uncorrelated and thus represent different aspects of research performance. Correlations with peer ratings or citation levels as a proxy, relate productivity or publication shares in the international subcategories to quality as perceived by international peers.

The findings are in agreement with language effects observed in citation-based measurements of research performance for university rankings (van Raan et al., 2011) and for the evaluation of national research performances (van Leeuwen et al, 2001). For the development of funding or evaluation models, these observations indicate that the different roles and characteristics of different publication media should be taken into account according to the model's aims.

The international dimension's role and its variability with country, language or discipline, can now more easily be studied, given the enhanced media coverage and author identification in international databases.

## Acknowledgement
Special thanks go to Arlette De Bruyn for the meticulous preparation of the team-based data and for her comments on earlier drafts of this paper.